\documentclass[aps,prd,reprint,nofootinbib,citeautoscript,floatfix]{revtex4-1}
\pdfoutput=1

\usepackage{mathtools}
\usepackage{amsmath, amssymb, amsthm, amsbsy, bbm}

\usepackage[percent]{overpic}
\usepackage{xcolor}
\usepackage[utf8]{inputenc}
\usepackage{enumitem}
\usepackage{hyperref}
\usepackage{graphicx}

\usepackage{subfig}

\usepackage[normalem]{ulem}
\usepackage{etoolbox}
\usepackage{cancel}
\usepackage{todonotes}

\newcommand{\bra}[1]{\langle #1 |}
\newcommand{\ket}[1]{| #1 \rangle}
\newcommand{\braket}[2]{\langle #1 \vphantom{#2}|
   #2 \vphantom{#1} \rangle}
\newcommand{\bracket}[3]{\langle #1 \vphantom{#2#3}|
  #2 | #3 \vphantom{#1#2} \rangle}
\newcommand{\ketbra}[2]{|#1\rangle \langle #2 |}

\newcommand{\be}{\begin{equation}}
\newcommand{\ee}{\end{equation}}
\def\({\left(}
\def\){\right)}
\def\[{\left[}
\def\]{\right]}

\theoremstyle{remark}

\theoremstyle{definition}

\newcommand{\calH}{\mathcal{H}}

\newcommand{\calO}{\mathcal{O}}

\newcommand{\He}{{\calH _{E}}}

\usepackage{cancel}

\newcommand{\tr}[1]{\mathrm{tr}\left[ #1 \right]}
\newcommand{\trr}[2]{\mathrm{tr}_{#1}\left[ #2 \right]} 

\newtoggle{editing}
\toggletrue{editing}

\iftoggle{editing}{%
	\setlength{\marginparwidth}{2cm}

	\newcommand{\jason}[1]{\textbf{\color{blue}[#1]}}
	\newcommand{\jamie}[1]{\textbf{\color{red}[#1]}}
	\newcommand{\moshe}[1]{\textbf{\color{green}[#1]}}
	\newcommand{\david}[1]{\textbf{\color{purple}[#1]}}
	
}{%
	\newcommand{\jason}[1]{\textbf{\color{blue}[]}}
	\newcommand{\jamie}[1]{\textbf{\color{red}[]}}
	\newcommand{\moshe}[1]{\textbf{\color{green}[]}}
	\newcommand{\david}[1]{\textbf{\color{purple}[]}}
	
}

\begin{document}

\title{Eigenstate Thermalization and Disorder Averaging in Gravity}

\author{Jason Pollack}
\email{jpollack@phas.ubc.ca}
\author{Moshe Rozali}
\email{rozali@phas.ubc.ca}
\author{James Sully} 
\email{sully@phas.ubc.ca}
\author{David Wakeham}
\email{daw@phas.ubc.ca}
\affiliation{Department of Physics and Astronomy, University of British Columbia, Vancouver, BC V6T 1Z1, Canada}

\begin{abstract}
Naively, a resolution of the black hole information paradox appears to involve microscopic details of a theory of quantum gravity.
However, recent work \cite{Penington:2019npb,Almheiri:2019psf,Almheiri:2019hni,Penington:2019kki,Almheiri:2019qdq} has argued that a unitary Page curve can be recovered by including novel replica instantons in the gravitational path integral.
Moreover, replica instantons seem to rely on disorder averaging the microscopic theory, without a definite connection to a single, underlying unitary quantum system. 
In this letter, we show that disorder averaging and replica instantons emerge naturally from a gravitational effective theory built out of typical microscopic states. 
We relate replica instantons to a moment expansion of the simple operators appearing in the Eigenstate Thermalization Hypothesis, describe Feynman rules for computing the moments, and find an elegant microcanonical description of replica instantons in terms of wormholes and Euclidean black holes.
\end{abstract}

\maketitle

\section{Introduction and Summary}

Recent discussions of the black hole information paradox have led to significant progress in understanding information loss in semiclassical effective field theory \cite{Penington:2019npb,Almheiri:2019psf,Almheiri:2019hni}. 
  Surprisingly, they have also shed light on how the semiclassical calculation may be rectified in order to be consistent with unitarity, without appealing directly to an underlying microscopic theory.\footnote{For related discussions see also \cite{Akers:2019nfi,Rozali:2019day,Almheiri:2019yqk,Almheiri:2019psy}.}

In these recent discussions, the inclusion of {\it replica instantons} is central 
to maintaining consistency
with unitary evolution \cite{Penington:2019kki,Almheiri:2019qdq}. These are Euclidean configurations which contribute to correlations between several copies of the theory, represented by distinct asymptotically AdS boundaries. A single unitary boundary theory with fixed couplings can have no such correlations, so gravitational calculations involving connected multi-boundary correlators are naturally interpreted in the context of an ensemble of theories. 
The goal of this letter is to clarify the origin of this statistical description.

The statistical description of a {\it single} quantum theory is familiar in the context of the Eigenstate Thermalization Hypothesis (ETH) \cite{PhysRevA.43.2046,PhysRevE.50.888,SREDNICKI_1995,Srednicki1999TheAT}. The basic idea is that a closed (isolated) chaotic many-body system, when probed only with simple (macroscopic) operators, looks for all intents and purposes thermal. This replaces the ideas of Gibbsian ensembles, or couplings to external heat baths, as a foundation for quantum statistical mechanics.  We are still discussing a single quantum system, evolving unitarily in a pure state: the effective coarse-graining comes from our limitations in gathering information about the system. 

Specifically, the matrix elements of a collection of simple operators $\{\calO_a\}$ in energy eigenstates $\{\ket{E_\ell}\}$ can be described in a chaotic system (without any simple conserved quantities apart from the energy) as
\begin{equation}\label{eq:ETH-correlator}
\langle E_i | \calO_a |E_j \rangle = f^{(a)}_{1}(E)\delta_{ij}+ e^{-S/2}f^{(a)}_{2}(E,\Delta E) R^{(a)}_{ij} \, .
\end{equation}
In a given theory, the variances of the matrix elements $R^{(a)}_{ij}$ are a fixed set of $O(1)$ numbers. 
If we lack sufficient information to distinguish a specific state, we can effectively replace the matrix elements by random variables that have the correct statistics. 

In the statistical description, to leading order the $R_{ij}$ are independent Gaussian random variables. 
We emphasize that using this statistical description does not necessarily mean that we are considering an ensemble of theories or coupling our system to an external bath; here, rather, we are interested in the properties of \emph{typical} states in a single theory.

In this letter, we argue that the correct objective for the semiclassical saddle-point expansion of low-energy effective field theory is the reproduction of the 
correlation functions of simple operators in typical microscopic states. 
These correlators are well-described by the 
ETH and we write down effective partition functions that generate their moments. 

We derive a set of `Feynman rules' for diagrammatically computing the partition functions for the moments of correlators.
And we show that, in holographic theories, the partition functions and their diagrammatic expansion may be understood in terms of gravitational path integrals. 
In the path-integral description, the higher moments are macroscopic quantities closely related to the replica instantons of \cite{Penington:2019kki,Almheiri:2019qdq}. 
Making a coarse assumption of chaos in the microcanonical ensemble, we find a particularly simple description of the higher moments and of replica instantons in terms of familiar Euclidean black holes connected by `wormholes'.






\section{Ensembles, quantum chaos, and the ETH}\label{sec:micro-physics-EFT}

In this section, we expand on the ETH and its relationship to low-energy effective field theory. 

Consider a \emph{microscopic} Hilbert space $\calH$ for a theory with a gravitational description. 
We are ultimately interested in computing quantities related to the physics of this quantum system. 
For concreteness, we take the theory to be a conformal field theory (CFT), in any dimension, with large central charge $c$. 
Within the microscopic Hilbert space, let us concentrate on the subspace of states within some microcanonical energy
window of width $\delta E$ about energy $E$, denoted $\He$.\footnote{Here we are considering the CFT quantized on a sphere of radius $R$, where the energy is simply related to the conformal dimension $\Delta$ by $E \sim  R \Delta$.} 
We will consider sufficiently high energies $E$ so that the microcanonical Hilbert space has dimension $\exp[S(E,\delta E)] \sim c$.
For the remainder of this letter, we suppress any dependence on $\delta E$. 

In our microscopic theory, 
we are naturally interested in correlators and transition amplitudes for simple operators and states within the window, e.g. 
\begin{equation}
    \braket{\psi_i}{\psi_j} \, , \; \bra{\psi_i} \calO_a \ket{\psi_i} \, ,  \; \bra{\psi_i} \calO_a \ket{\psi_j}\, \, \ldots 
\end{equation}
for states $\ket{\psi_i},\ket{\psi_j} \in \He$ and some appropriately-chosen collection of simple operators $\calO_a$.
We usually think of the operators as 
`small' products of local operators, each with $\Delta \sim O(1)$. In the Heisenberg picture this means we also exclude operators evolved for too long in time.\footnote{In the Schr\"{o}dinger picture, an initial state which is typical in the microcanonical window may become atypical after exponentially long times due to quantum ergodicity.} The precise choice we make is unimportant to the argument of this letter.

To design an effective field theory we also require a specification of a state or of a distribution over states. 
A coarse-graining over states reflects uncertainty in determining the true microscopic state using our simple low-energy operators, as well as uncertainty in how the original microscopic state was produced.

What is the correct ensemble of states to study when we coarse-grain? 
Our goal in this letter is not necessarily to solve this problem exactly, but to explore the consequences in effective field theory. 
Nevertheless, to be concrete, we will attempt to build a sensible distribution at the coarsest level. 

Although energy is a conserved charge, restricting our effective field theory to simple operators for finite times limits the ability of low-energy observers to probe the exact energy of microstates. 
Only after times exponentially large in the entropy $S$ can operators probe the energy splittings in the microcanonical window.\footnote{Crudely, a typical energy splitting is order $\delta E/e^S$, which requires time $\delta t \sim \hbar e^S/\delta E$ to probe.}
Were there other conserved charges accessible to our simple operators $\calO_a$, these could be measured to further refine the microcanonical ensemble into sub-ensembles conditioned on the measurement of these charges, just as grand ensembles are used in statistical mechanics. 

We will assume, instead, that our system is chaotic. 
For the purposes of this letter, we identify chaos by the fact that no such charges are measurable by our simple operators. 
As a result, typical states relevant to physical processes are indistinguishable from those drawn at random from $\He$ by applying a Haar-random unitary in $L(\He)$ to a reference state $\ket{\psi_0}\in \He$.\footnote{That is, we are working in the Circular Unitary Ensemble.}
In this case, for typical states $\ket{\psi_i},\ket{\psi_j} \in \He$ we expect
\begin{equation}\label{eq:CLT-correlator}
    \bracket{\psi_i}{\calO_a}{\psi_j} = f_1^{(a)}(E)\delta_{ij} + e^{-S(E)/2}f_2^{(a)}(E) R^{(a)}_{ij} \, 
\end{equation}
simply by the central limit theorem. Exactly as in the discussion of the ETH above, $f_1^{(a)}(E)$ is the average of an operator's microcanonical eigenvalues and $f_2^{(a)}(E)^2$ its variance, while
$R^{(a)}_{ij}$ has the statistics of a matrix of iid random complex numbers with zero mean and unit variance.
Note that although the entries for an individual matrix are iid, matrices for different operators will typically exhibit correlations, with a smooth covariance
\begin{equation}
    \overline{R^{(a)}_{ij}R^{(b)}_{kl}} \equiv \delta_{il}\delta_{jk} \sigma_2^{(ab)} \, .
\end{equation}
As we will show below, this covariance is simply related to microcanonical operator traces.

Our assumption that states cannot be distinguished within the microcanonical ensemble by simple operators is in essence a restatement of the ETH, as in Eq.\ \eqref{eq:ETH-correlator}. 
There, the energy eigenstates themselves, as probed by simple operators, look like typical microcanonical states. 
Note that in the ETH, the functions $f_2^{(a)}$, and higher moments $f_n^{(a)}$, depend on the energy differences $\Delta E = E_i-E_j$, as well as the average energy $E$. 
However, if our energy window is narrower   than the Thouless energy this dependence disappears \cite{DAlessio:2016rwt}. 
We will limit ourselves to this regime in the following. 

In summary, we will build an effective field theory to describe the typical, expected value of correlators or, with even better precision, sufficiently large sums or averages over microscopic states, such as
\begin{equation}
   \overline{\bra{\psi_i} \calO_a \ket{\psi_j}} \, , \; \sum^{e^K}_{i,j} \bra{\psi_i} \calO_a \ket{\psi_j} \, . \ldots  
\end{equation}
The effective field theory should describe {\it statistical} properties of this set of correlators, that is, the moments of correlation functions $f_n^{(a)}, \, \sigma_n^{(a_1 \ldots a_n)}$. 
We will see that this results in apparent ``disorder averaging" in the effective description.
For the holographic theories under consideration, the typical correlators will be simply determined by semiclassical gravitational saddles.

\section{Generating functions for mean correlators} 

We start with the simplest case: the use of our effective field theory to calculate the averaged correlators 
\begin{align}
    \overline{\bracket{\psi_i}{\calO_a}{\psi_j}}  & = \delta_{i j} e^{-S}\trr{\He}{\calO_a} =\delta_{i j} f_1^{(a)} (E) \, .
\end{align}
To summarize those observables, we can write a generating function for the microcanonical mean values as
\begin{align}\label{eq:micro-generating-lo}
    Z^{(1)}_{ij}(E,J_a) \equiv \sum_a J_a  \overline{\bracket{\psi_i}{\calO_a}{\psi_j}}  \nonumber \\
    = \delta_{ij} e^{-S} Z^{(1)}(E,J_a)   \, ,
\end{align}
where
\begin{align}
    Z^{(1)}(E,J_a) &\equiv  \sum_a J_a \trr{\He}{\calO_a}\, .
\end{align}
By ``generating function,'' we mean as usual that derivatives with respect to sources give expectations:
\begin{equation}
\frac{\partial Z_{ij}^{(1)}}{\partial J_a}\bigg|_{J=0} = \overline{\bra{\psi_i}\calO_a\ket{\psi_j}}.
\end{equation}
We are implicitly including $\calO_0 = \mathbb{I}$ in the sum, with $J_0 = 1$ fixed.
We make this choice for all microcanonical generating functions in the remainder of this letter.\footnote{The simple operators $\{\calO_a\}$ do not form an algebra, since arbitrary higher-point correlators should not be part of our effective theory. Our generating functions are thus defined to give expectations in the linear span of $\{\calO_a\}$ only.}

\subsection{Feynman rules for the mean partition function}\label{mean-feynman-rules}

Let us begin here to introduce some `Feynman rules' to compute the mean partition function. 
These follow from more standard diagrammatics for unitary integrals (eg. \cite{Brouwer1996}), but we have chosen a notation particularly suited to the case at hand.  
We will extend these rules in subsequent sections as we explain how to obtain the higher moments. 

We indicate a correlator $\bracket{\psi_i}{\calO}{\psi_j}$ by a vertex, associated with a numerical factor $e^{-S}$:
\begin{center}
    \includegraphics[scale=0.3]{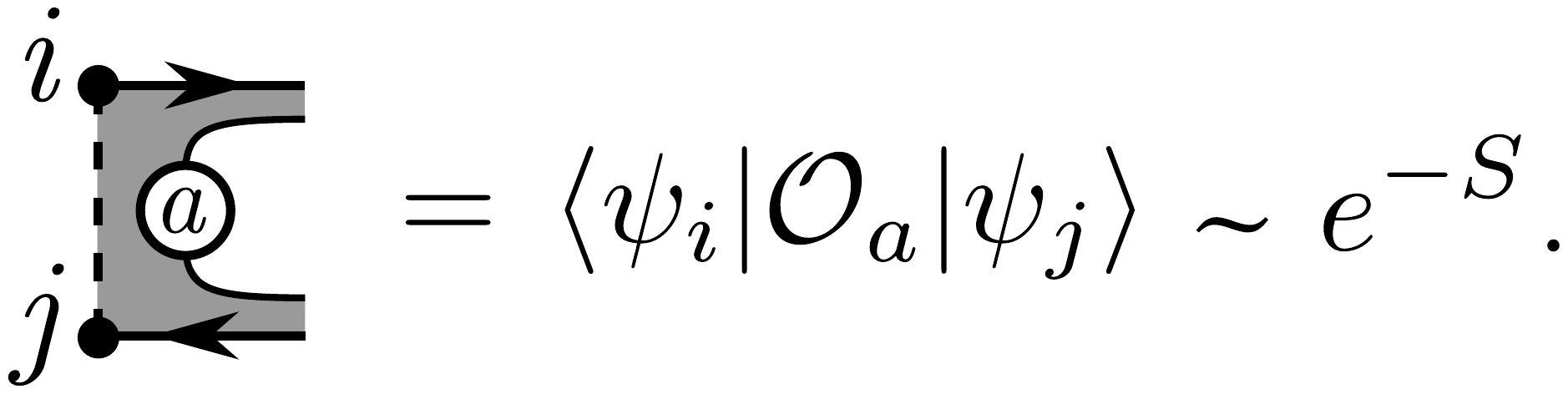}
\end{center}
The outer lines carry an index of the state, with an arrow indicating whether it is a `bra' or a `ket'. 
The inner lines carry the index structure of the operator that is inserted. 
When the indices $i, j$ of the outer lines are equal, they can be contracted (and similarly for indices $m, n$ of the operator trace):
\begin{center}
    \includegraphics[scale=0.3]{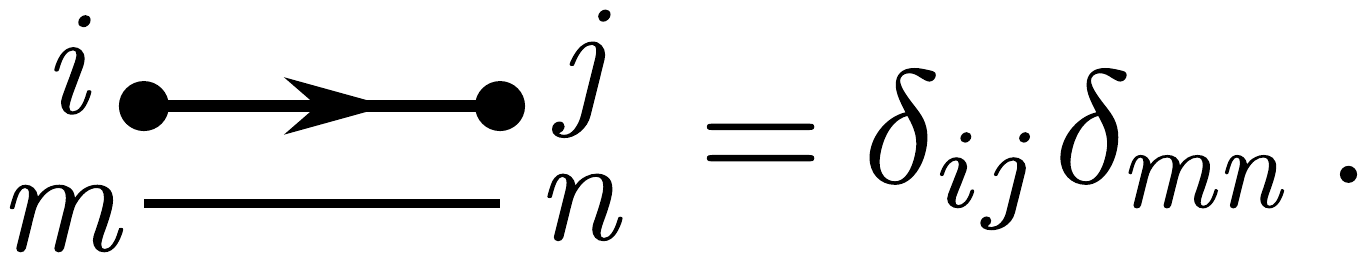}
\end{center}

When contracted, they form a geometry with the topology of a disk, inside of which there is a loop indicating the contraction of the indices of the inserted operator.
Thus, our mean partition function is computed by the diagram
\begin{center}
    \includegraphics[scale=0.3]{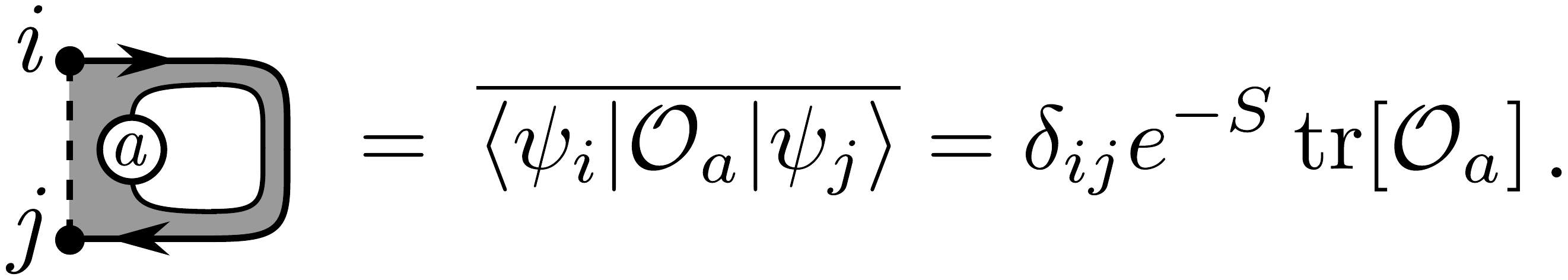}
\end{center}

\subsection{Example: the microcanonical partition function}

The definition \eqref{eq:micro-generating-lo} immediately leads to a simple expression for the microcanonical partition function by summing over $e^S$ random states, converging to their mean:
\begin{align}
    Z_{\overline{\text{CFT}}} (E,J_a) &\equiv \sum_a J_a \sum_i^{e^S} \bracket{\psi_i}{\calO_a}{\psi_i} \nonumber \\
    &\approx Z^{(1)}(E,J_a) \, .
    \label{eq:Z1}
\end{align}
Diagrammatically, we can write
\begin{center}
    \includegraphics[scale=0.3]{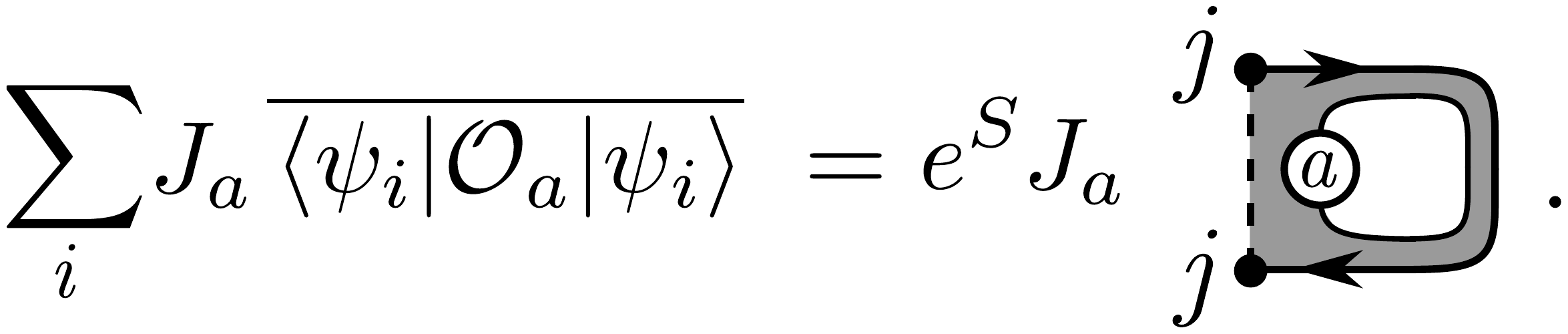}
\end{center}

It is often more standard to consider a canonical generating function, especially for gravitational effective theories.  
Then we have the coarse-grained, thermal CFT partition function
\begin{align}
    Z_{\overline{\text{CFT}}}(\beta,J_a) &\equiv  \sum_{E= E_0 + n\delta E} e^{ - \beta E} Z_{\overline{\text{CFT}}} (E,J_a) \nonumber\\
    &\approx \int dE \,\rho(E) e^{ - \beta E} \sum_a {J_a}  f_1^{(a)}(E) \, ,
\end{align}
where we have approximated the sum as an integral when $E \sim c$ is large and $\delta E / E \sim c^{-\alpha}$,  $1/2 < \alpha < 1$. 
Here $\rho(E) = \exp (S) /\delta E $ is the density of states. 

\subsection{The gravitational description}

The partition function for a CFT at finite temperature is prepared by a path integral on $S_d \times S_1$. 
When a bulk dual exists, the gravitational picture is well-known (see e.g.\ \cite{Witten:1998zw,Horowitz:1999jd,Maldacena:2001kr}). 
At high temperatures, the leading semiclassical saddle to the bulk gravitational partition function with boundary $S_d \times S_1$ is a Euclidean black hole. 
One can compute simple bulk correlation functions in this background, and find that their boundary limit matches the leading-order thermal CFT correlation function \cite{KeskiVakkuri:1998nw,Maldacena:2001kr,Son:2002sd}.

Like our effective generating function, the bulk saddle is not sensitive to the exponentially small level splittings; this can seen most easily in the real-time analytic continuation, where bulk correlators continue to decay for all time without random noise at Heisenberg time scales and without quantum Poincar\'{e} recurrences \cite{Maldacena:2001kr,Barbon:2003aq,Birmingham:2002ph,Barbon:2004ce}. 

Note that the canonical picture is not essential to this story. 
With a little more effort, one can similarly find the bulk solution dual to the microcanonical partition function \cite{Marolf2018l}. 
As long as the microcanonical width scales as $O(1)< \delta E < O(c^{1/2})$, the projection of the bulk gravitational path integral onto a microcanonical band results in a single semiclassical bulk geometry. 

Thus, we can equate our Feynman diagram to a gravitational geometry:
\begin{center}
    \includegraphics[scale=0.3]{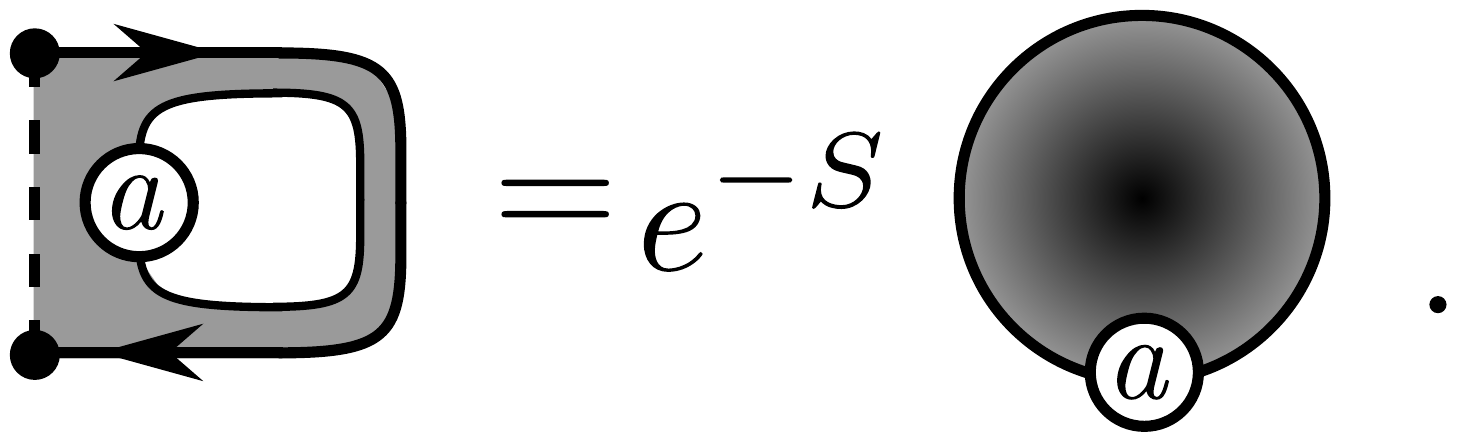}
\end{center}
with an equivalence of partition functions
\begin{equation}
    Z^{(1)}(E,J_a) = Z_{grav}(E,J_a) \approx Z_{grav}(\beta_E,J_a)\, .
\end{equation}
\section{Generating functions for second moments}

So far, we have only required that the saddles of our effective field theory describe the mean microcanonical value of simple correlation functions.
However, physical processes may probe slightly more fine-grained `mesoscopic' questions about the CFT, namely the quantities $f_n^{(a)}, \; \sigma_n^{(a_1 \ldots a_n)}$ defined in Sec.\ \ref{sec:micro-physics-EFT}. The simplest such quantities are the covariances of the distribution of $R^{(a)}_{ij}$. 
We can extract the covariances from quantities of the form
\begin{align}
    &\bra{\psi_i} \calO_a \ket{\psi_j}\bra{\psi_k} \calO_b \ket{\psi_l} \, ,
\end{align}
now involving two copies of the theory. 
Using 
Haar averages over the unitary group, one can check (Appendix \ref{app:unitary-integrals}) that
\begin{align}\label{eq:second-order-sums}
    & \overline{\bra{\psi_i} \calO_a \ket{\psi_j}\bra{\psi_k} \calO_b \ket{\psi_l}} \nonumber \\ 
    &\quad =  \delta_{ij}\delta_{kl} e^{-2S}  \left(1-\tfrac{\delta_{jk}}{e^S+1}\right)\tr{\calO_a}\tr{ \calO_b}    \nonumber\\
    &\quad \quad+  \delta_{jk}\delta_{li}e^{-2S} \left(1-\tfrac{\delta_{ij}}{e^S+1}\right) \tr{\calO_a \calO_b}
    \, ,  
\end{align}
where the traces are over $\He$ (we drop the subscripts when clear from context) and products of operators have indices contracted only in the microcanonical subspace. 

The first line in the above expression depends only on mean values.
At leading order, it is just the product of disconnected mean generating functions. 
However, it also receives an $e^{-S}$ correction.\footnote{Technically the correction is $\tfrac{1}{e^S +1}$, and so we might wish to think of this as re-summing an infinite number of $e^{-nS}$ contributions.} 
The second line is a connected contribution not derivable from the mean generating function. 
Like the first line, it also receives an $e^{-S}$ correction. 

We can write a generating function for the general second moment as
\begin{align}\label{eq:second-moment-generic}
    Z_{ij,kl}&(E,J_{1,a},J_{2,b})  \equiv \sum_{a, b} J_{1,a}J_{2,b}\overline{\bra{\psi_i}\calO_a\ket{\psi_j}\bra{\psi_k}\calO_b\ket{\psi_l}}
    \nonumber\\
    &=   Z^{(1)}_{ij}(E,J_{1,a})  Z^{(1)}_{ij}(E,J_{2,b})  \left(1-\tfrac{\delta_{jk}}{e^S+1}\right)    \nonumber\\
    &+  Z^{(2)}_{ij,kl}(E,J_{1,a},J_{2,b})  \left(1-\tfrac{\delta_{ij}}{e^S+1}\right)
\end{align}
for
\begin{align}
    Z^{(2)}_{ij,kl}(E,J_{1,a},J_{2,b}) &\equiv \delta_{jk}\delta_{li}e^{-2S}Z^{(2)}(E,J_{1,a},J_{2,b}) \nonumber \\ Z^{(2)}(E,J_{1,a},J_{2,b}) &\equiv \sum_{a,b} J_{1,a} J_{2,b}\tr{\calO_a \calO_b}\,. 
\end{align}

\subsection{Feynman rules for the second moment partition function}\label{sec:feynman-rules-second}

With the Feyman rules we introduced to compute the mean correlator, we can already compute the leading order contribution on each line of \eqref{eq:second-order-sums}:
\begin{center}
    \includegraphics[scale=0.3]{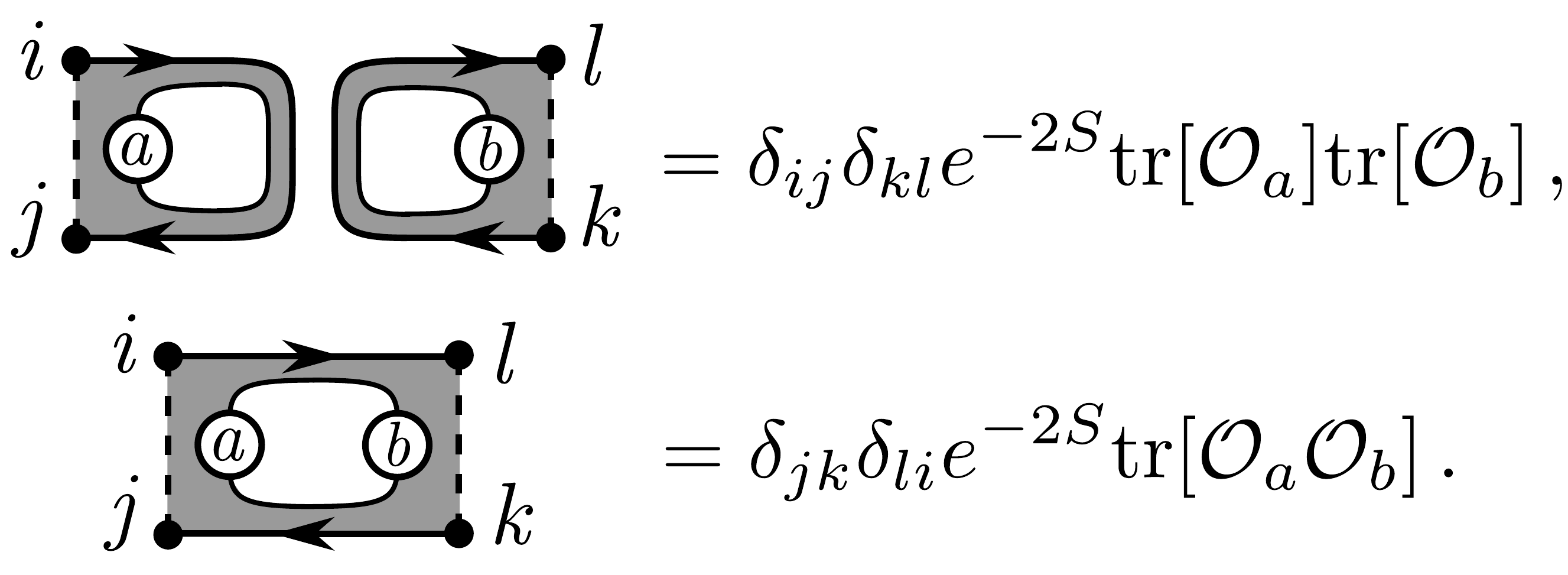}
\end{center}
To compute the subleading terms, we need to introduce a new vertex
which carries a power of $\frac{1}{e^{S}+1}$ and enforces that the state index passing along the lines it joins are equal:
\begin{center}
    \includegraphics[scale=0.3]{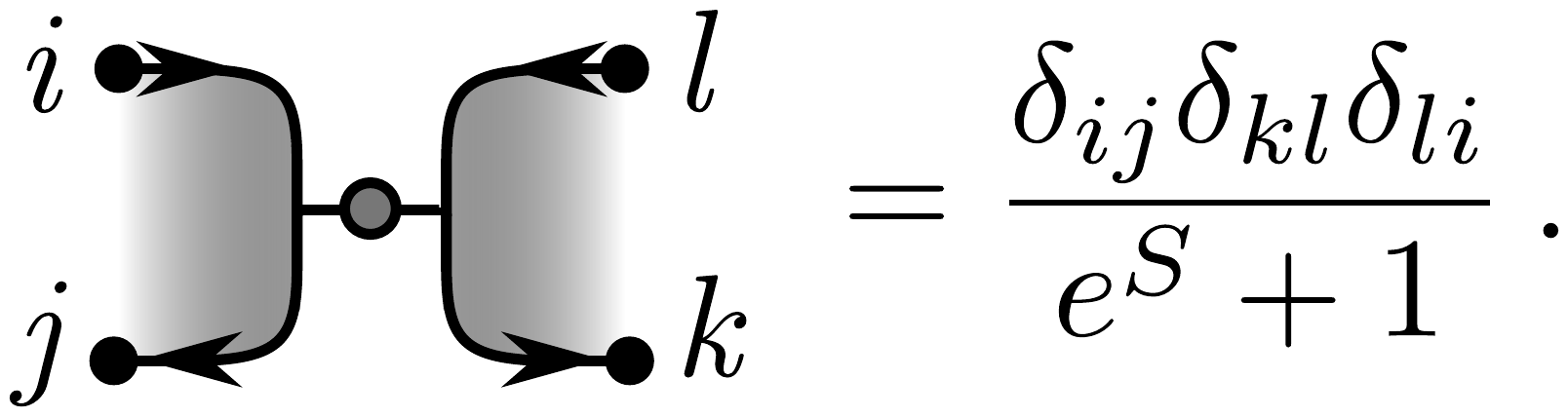}
\end{center}
We thus have Feynman diagrams for the corrections:
\begin{center}
    \includegraphics[scale=0.3]{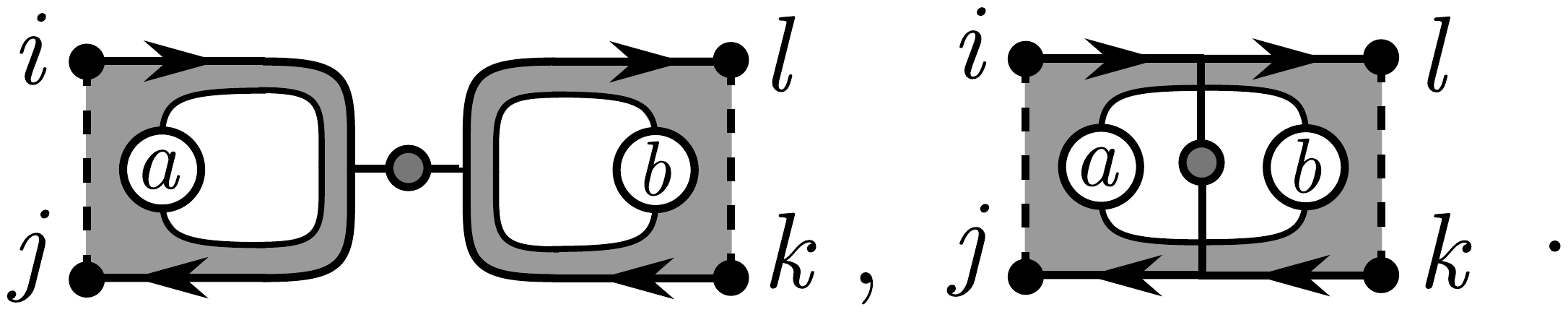}
\end{center}

\subsection{The gravitational description}\label{sec:second-moment-gravitational}

We already argued that the gravitational description of $Z^{(1)}$ is part of the standard AdS/CFT dictionary. 
Now the gravitational description of $ Z^{(2)}$ requires just a slight  elaboration. 
We seek a gravitational partition function to generate correlators of the form
\begin{equation}
    \trr{\He}{\calO_a \calO_b} \, .
\end{equation}
Recall that these are microcanonical traces where operator indices have also only been contracted within the microcanonical subspace. 

Just as before, where the microcanonical partition function was given at leading order by the microcanonical black hole saddle, the insertion of another simple operator in the trace does not shift to another saddle and we again can compute the correlator using the same bulk solution. 
And as the same energy runs between both operators, they must be equally spaced on opposite sides of the circular Euclidean-time boundary. 

Again, the reader may be more familiar with the canonical picture. Translating into the canonical language, we have the approximate identity (in the thermodynamic limit)
\begin{equation}
    \trr{\He}{\calO_a \calO_b} \approx \trr{\calH}{e^{-\beta_E H/2 }\calO_a e^{-\beta_EH/2 }\calO_b}\,.
\end{equation}
Here it is even more apparent that the operators are equally spaced on opposite sides of the thermal circle. 

Thus, we can write
\begin{align}
    Z^{(2)}(E,J_{1.a},J_{2,b}) &= Z_{\text{Grav}}(E,J_{1.a},J_{2,b}) \nonumber\\
    &\approx Z_{\mathrm{Grav}}(\beta_E,J_{a},J^{(1/2)}_{b}) \, ,
\end{align}
where $J^{(1/2)}$ is the source for the operator $e^{-\beta_E H/2} \calO_b e^{\beta_E H/2}$.

Our immediate take-away is that the Euclidean wormhole needed to compute $ Z^{(2)}$ is just the standard wormhole for the microcanonical (or thermal) black hole.

Furthermore, while we have no direct gravitational interpretation of the corrections to the leading terms, we have suggested that they might be thought of as topologically non-trivial wormholes that glue the geometries together:
\begin{center}
    \includegraphics[scale=0.3]{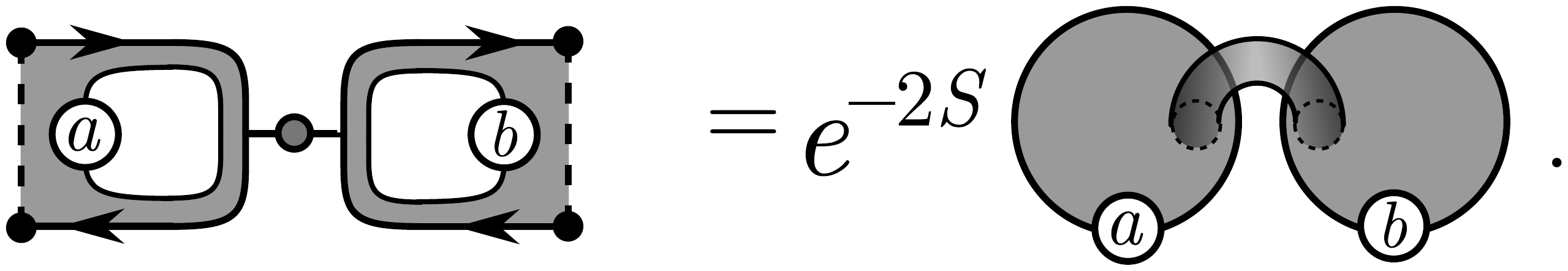}
\end{center}
See \cite{Penington:2019kki} for a related discussion of `handles' joining replica instantons.

\subsection{Example 1: Product of two CFTs}\label{sec:product-partition-functions}

We know that in the full microscopic theory, the microcanonical partition function for a CFT on $(S_d \times  S_1)^2 $ is just a square of the partition function on one copy.  
But while the sum over eigenstates factorizes as expected,
\begin{equation}
    \left( \sum_{i} \braket{E_i}{E_i}  \right) \left( \sum_{j} \braket{E_j}{E_j} \right) \, ,
\end{equation}
when we insert simple operators, the ETH tells us that the off-diagonal terms cancel to good approximation, leaving only the diagonal second moment:
\begin{align}
     \sum_{i} \bracket{E_i}{\calO_a}{E_i} & \sum_{j} \bracket{E_j}{\calO_b}{E_j} =  \nonumber \\
    &e^{2S} f_1^{(a)} f_1^{(b)}  + \sigma_2^{(ab)}f_2^{(a)} f_2^{(b)} + \ldots\:.
\end{align}
Thus, the factorization of the partition function is actually misleading in terms of the non-apparent factorization of correlators in this expansion.

We can nevertheless generate the connected contribution to the partition function squared from a partition function that is itself connected. 
We may replace the sum over eigenstates with an equal number of typical states
\begin{equation}
    \sum_{i} \ketbra{E_i}{E_i} \rightarrow     \sum_{i} \ketbra{\psi_i}{\psi_i},
\end{equation}
since summation of large numbers of typical states is sufficient to calculate microcanonical averages. 
This leads us to the partition function we considered in the mean case, $Z_{\overline{\text{CFT}}}(E,J_{1,a})$. 
Using \eqref{eq:Z1} and \eqref{eq:second-moment-generic}, we see that 
\begin{align}
    Z_{\overline{\text{CFT}}}&(E,J_{1,a})   Z_{\overline{\text{CFT}}}(E,J_{2,b})\ \approx  \nonumber\\   
    & Z^{(1)}(E,J_{1,a})   Z^{(1)}(E,J_{2,b}) (1- \tfrac{1}{e^{S}(e^S +1)}) \nonumber \\ 
    &\qquad  + e^{-S} Z^{(2)}(E,J_{1.a},J_{2,b}) (1- \tfrac{1}{e^S +1}) \:.
\end{align}
We can write this diagrammatically as
\begin{center}
    \includegraphics[scale=0.22]{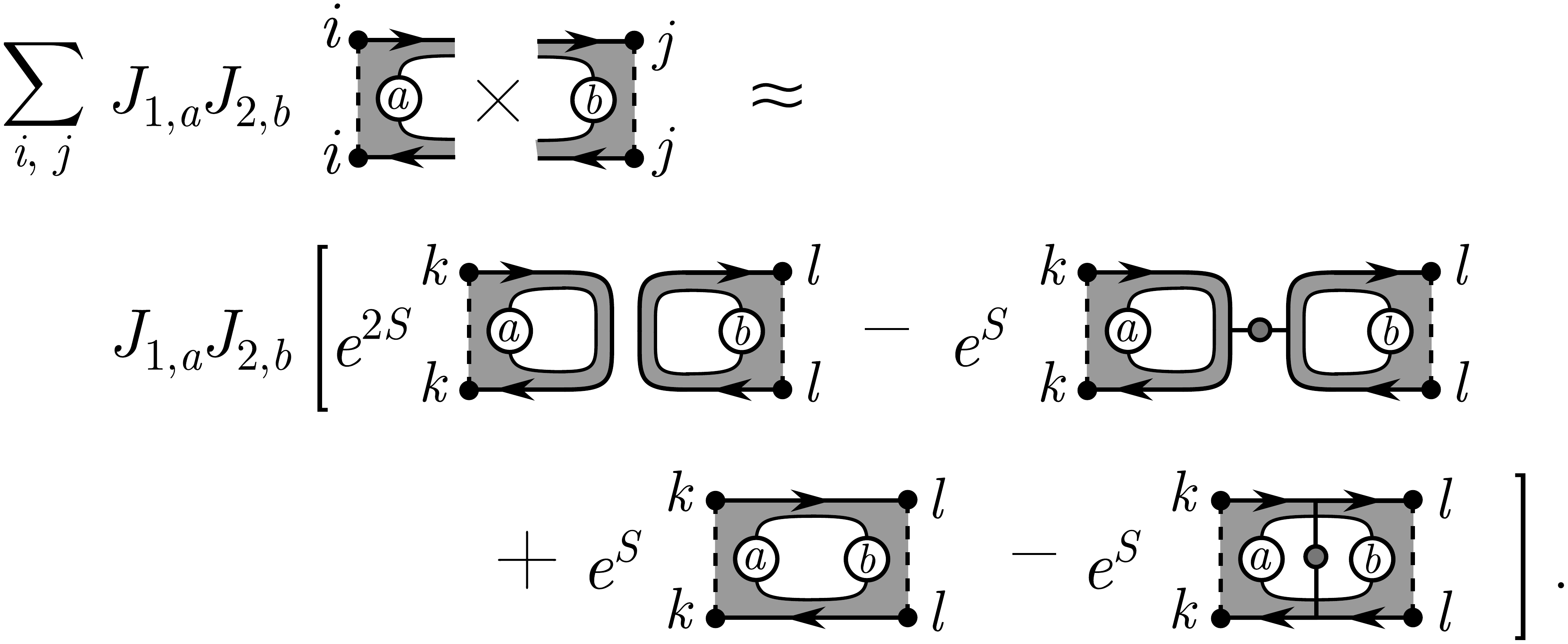}
\end{center}

\subsection{Example 2: Second R\'{e}nyi partition function}\label{sec:second-renyi}

To calculate the second R\'{e}nyi entropy of the matrix
\begin{equation}
    \rho_{K}  \equiv e^{-K} \sum_{i}^{e^K} \ketbra{\psi_i}{\psi_i},
\end{equation}
we need to compute $\tr{\rho_K^2}$. 
We can view this more generally as a partition function of the form
\begin{equation}
    Z_{K,2} (E, J_{1,a}, J_{2,b})  = \sum_{a,b}  J_{1,a} J_{2,b} \tr{\rho_K \calO_a \rho_K \calO_b  } \, .
\end{equation}
From Eq.\ \eqref{eq:second-moment-generic}, this is
\begin{align}
   Z_{K,2} & (E, J_{1,a}, J_{2,b}) \\ 
   &\approx e^{-2S-K} Z^{(1)}(E,J_{1,a})Z^{(1)}(E,J_{2,b}) (1- \tfrac{1}{e^S +1})  \nonumber \\ 
   & \qquad+ e^{-2S} Z^{(2)}(E,J_{1.a},J_{2,b}) (1- \tfrac{1}{e^{K}(e^S +1)}) \,. \nonumber
\end{align}
Note that, while the connected correlator was suppressed in the squared CFT partition function, here it can grow to equal size when $K \sim S$.

To calculate the second R\'{e}nyi entropy, $S_2(\rho_K) = - \log Z_{K,2}$, we take $\calO_{a,b} = \mathbb{I}$ and find
\begin{equation}
    S_2(\rho_K) = -\log\big(e^{-K}+ e^{-S}-e^{-S -K}\big) \, .
\end{equation}

\section{Higher moments}

We can similarly compute higher moments of our correlation functions. 
Using Haar averages, one can show that
\begin{align}\label{eq:higher-moment-average}
    &\overline{\prod_{m=1}^{n} \bracket{\psi_{i_m}}{\calO_{a_m}}{\psi_{j_m}}} = \nonumber\\
   &e^{-nS}  N\big(\vec{i}\big) \sum_{\sigma \in S_n} \prod_m \delta_{i_m,j_{\sigma(m)}}\delta^{k_m}_{l_{\sigma(m)}} \left(\calO_{a_m}\right)_{k_m l_m} \, ,
\end{align}
where $\sigma$ is a permutation. 
The normalization $N\big(\vec{i}\big)$ and details of the proof can be found in Appendix \ref{app:unitary-integrals}. 
In short: each permutation in \eqref{eq:higher-moment-average} is described at leading order by one of our Feynman diagrams showing how identical states are joined together to form traces of the operators. For example, the fourth moment contains a term 
\begin{center}
    \includegraphics[scale=0.29]{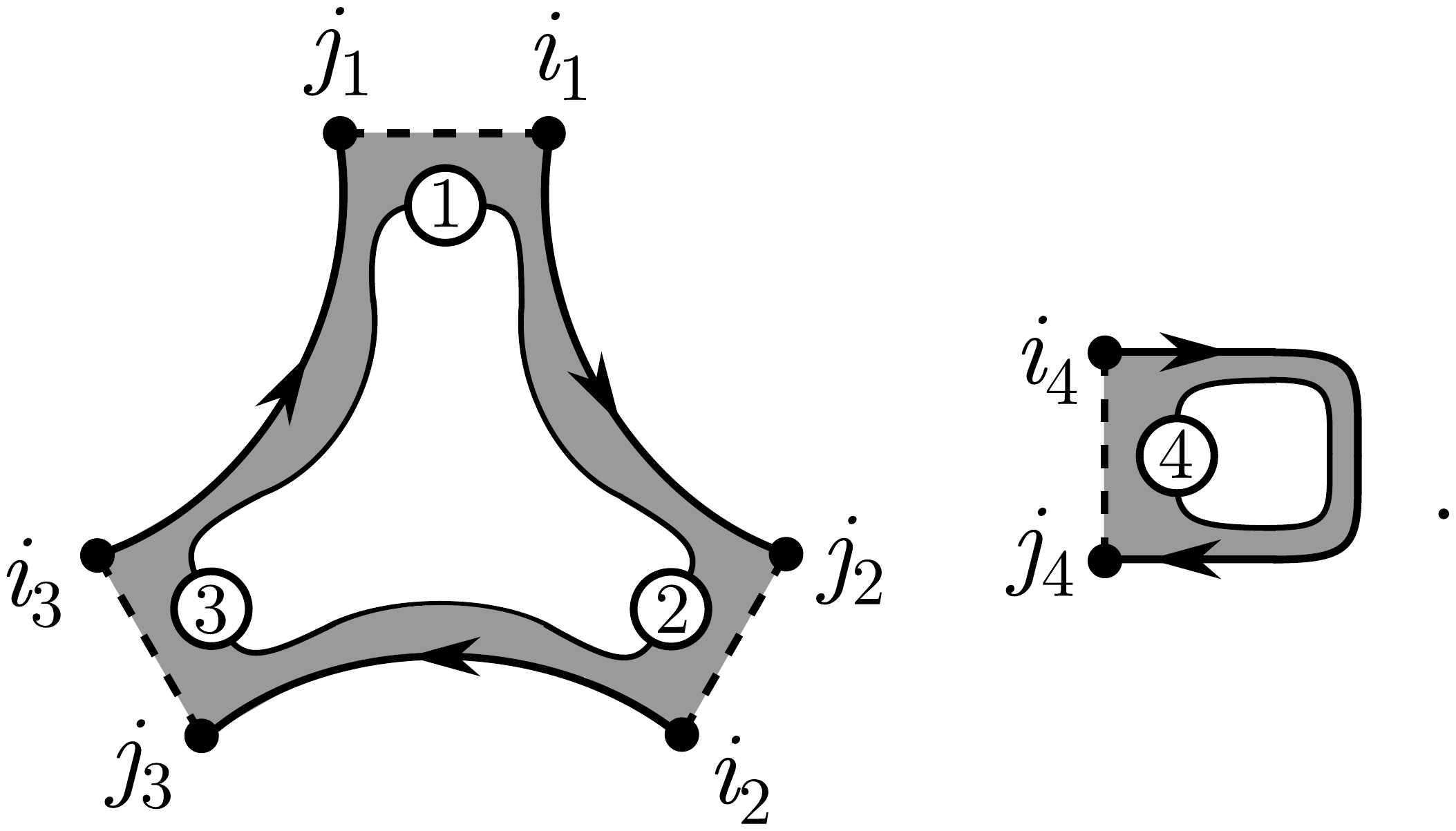}
\end{center}

To compute the normalization, we must resum an infinite series of tree-level contributions from vertices that link the boundaries together. 
We already showed in Section \ref{sec:feynman-rules-second} how to compute the leading order corrections,
\begin{equation}
    N(\vec{i}) = 1 - \sum_{\substack{\text{pairs}\\(i_m i_n)}}\frac{\delta_{i_m,i_n}}{e^S+1}+ \ldots \, ,
\end{equation}
in terms of vertices that join two boundaries together. 
Further corrections take the form 
\begin{equation}
    \sum_{k=3}^\infty \sum_{\substack{\text{sets} \\ (i_{m_1} \cdots i_{m_k}) }} c_k(S) \frac{\delta^{(k)}_{(i_{m_1} \cdots i_{m_k})}}{e^{(k-1)S}}, 
\end{equation}
where
\begin{equation}
     \delta^{(k)}_{(i_{m_1} \cdots i_{m_k})} = 
     \begin{cases}
    1 & \text{if $i_{m_a} = i_{m_b}$ $\forall$ $a,b$}\\
    0 & \text{otherwise}.
  \end{cases}
\end{equation}
We can represent each higher-order correction as a $k$-point vertex with a Feynman diagram 
\begin{center}
    \includegraphics[scale=0.27]{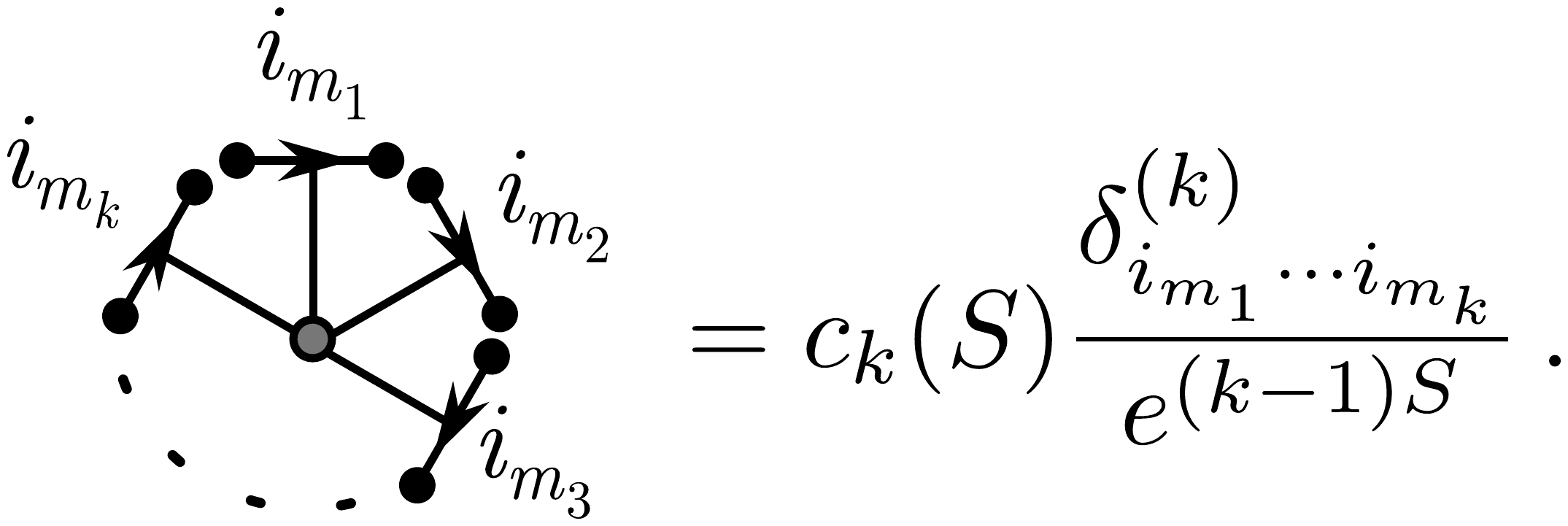}
\end{center}
The terms $c_k(S)$ are $O(1)$ combinatorial objects obeying a simple recursion relation.

\subsection{The gravitational description}

As each leading-order term needed to compute the higher moments is just a product of microcanonical traces, it will be calculated by the same gravitational saddles as the microcanonical black hole of energy $E$ without operator insertions. 
For each trace with $p$ operators inserted, we space them equally in order of the trace around the Euclidean time circle. 
The canonical picture is likewise simple. 
The canonical generating function for the trace with $p$ operators is just given by 
\begin{equation}
    Z_{\mathrm{Grav}}(\beta_E,\lbrace J^{(m/p)}_{a_m}\rbrace_{m=0}^{p-1})\,,
\end{equation}
where $J^{(m/p)}_{a_m}$ is a source for the operator $e^{-\beta_E H m/p} \calO_{a_m} e^{\beta_E H m/p}$.

As in the case of the second moment, we can view the subleading corrections to each trace as an infinite series of $k$-boundary wormholes joining the true gravitational geometries together, for instance
\begin{center}
    \includegraphics[scale=0.29]{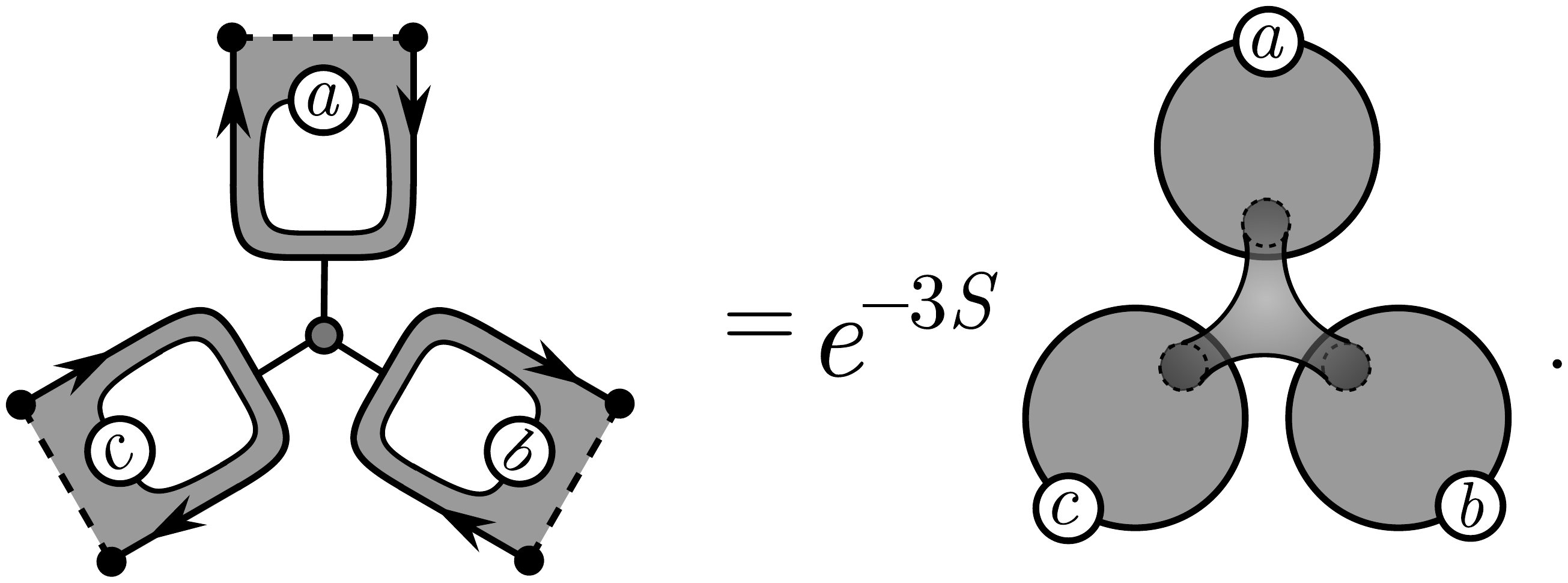}
\end{center}
Again, we have no true gravitational solution dual to these corrections, and view this as a heuristic description to motivate further work. 


\subsection{Example: higher R\'{e}nyi partition functions}\label{sec:higher-renyi}

As an example, let us consider the particular case of $Z_{K,n} =\tr{\rho_K^n}$. 
Again, we will think of this as a generating function for correlators with simple operators inserted between each $\rho_K$. 
Let us simplify our notation by inserting a fixed sequence of operators $\lbrace \calO_m \rbrace$, and expand the trace as 
\begin{equation}
    Z_{K,n} = e^{-nK}\sum_{\lbrace{i_{p}\rbrace}_{p=1}^n}^{e^K} \prod_{m=1}^n  \bracket{\psi_{i_m}}{\calO_m}{\psi_{i_{m+1}}} \, .
\end{equation}
For simplicity, we will ignore the `wormhole' corrections and compute only the leading term for each trace type.
By use of \eqref{eq:higher-moment-average}, we can rewrite this as
\begin{equation}
Z_{K,n} \approx \sum_{\sigma \in S_n} e^{-[nS+d(\sigma,\sigma_n)] K} \prod_{i=1}^{c(\sigma)}\tr{\prod_{j=1}^{d_i} \calO_{C_{ij}}} \, ,
\end{equation}
where $\sigma_n = \left( 1 \,2 \, \ldots \, n \right)$ and $d(\cdot,\cdot)$ is the Cayley distance between the permutations, measuring the minimal number of transpositions to change one permutation to the other. 
Each such transposition $(j \, k)$ encodes an equality constraint $\psi_{i_j} = \psi_{i_k}$ that allows us to contract the corresponding indices and reduces the number of contributing terms in the sum by $e^{-K}$. 
We have decomposed each permutation uniquely as a product of disjoint cycles $\sigma = C_1\cdot C_2 \cdots C_{c(\sigma)}$, where here we are also counting possibly-trivial cycles. 
Each trace corresponds to a cycle $C_i= \left( C_{i1} \,\ldots C_{i d_i}\right)$ of length $d_i$. 
See Appendix \ref{app:unitary-integrals} for further details. 

For $\calO_i = \mathbb{I}$, the above expression simplifies to 
\begin{equation}
Z_{K,n} \approx \sum_{\sigma \in S_n} e^{-d(\sigma,\sigma_n) K  - d(\sigma,\text{Id}) S} \, ,
\end{equation}
allowing one to explicitly compute the `leading trace' contribution to the R\'{e}nyi entropies for any $n$. 

We can also rewrite this partition function in terms of canonical gravitational partition functions (adding back in the dependence on sources) as
\begin{align}
    &Z_{K,n}(E, \lbrace J_{m,a} \rbrace ) \approx  \\
    &\sum_{\sigma \in S_n} e^{-d(\sigma,\sigma_n)K  - n S} \prod_{i=1}^{c(\sigma)}Z_{\mathrm{Grav}}(\beta_E , \lbrace J^{(j-1)/d_i}_{C_{ij},a} \rbrace_{j=1}^{d_i})\,. \nonumber
\end{align}

\section{Discussion}
\begin{figure}
    \centering
    \includegraphics[scale=0.35]{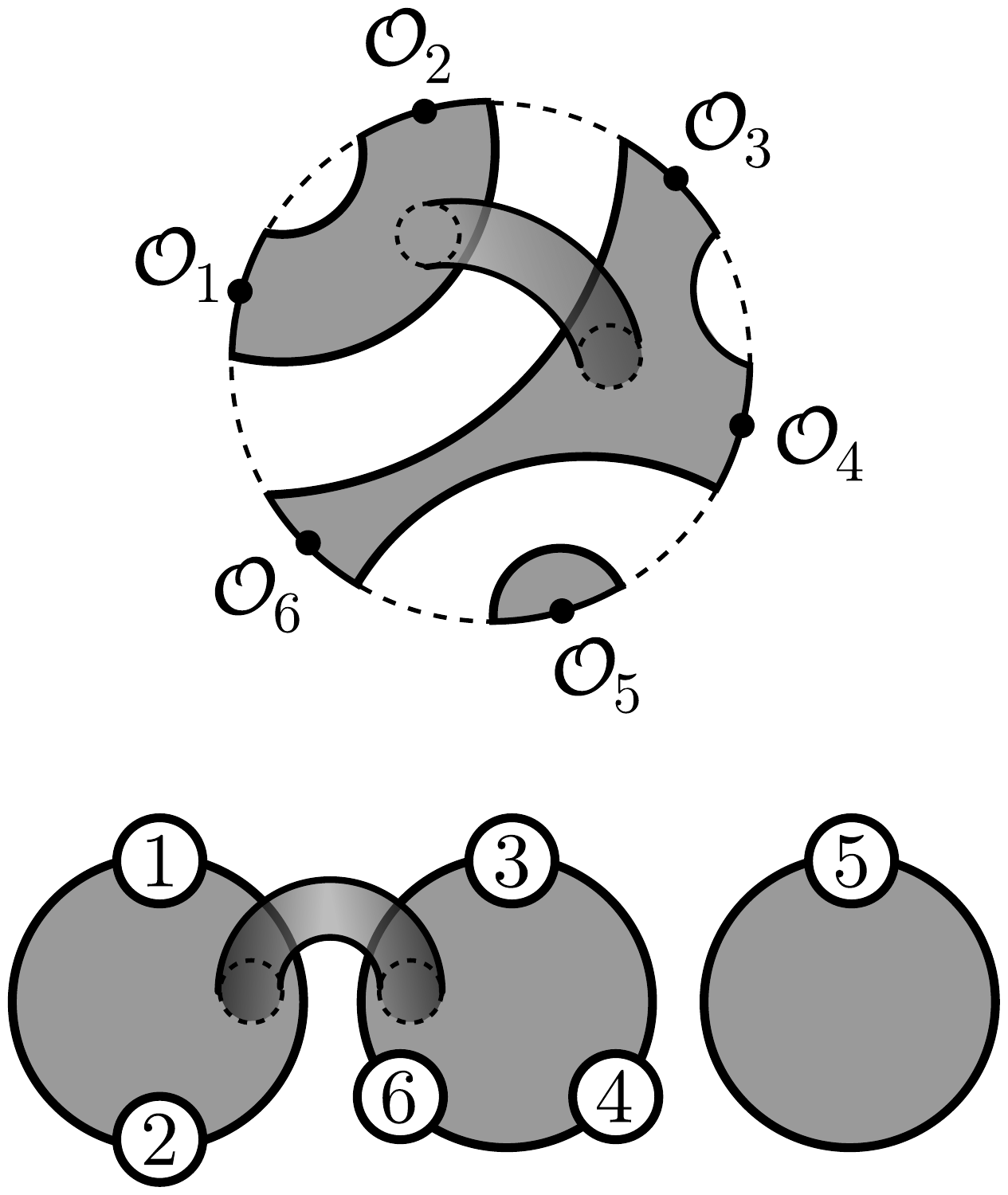}
    \caption{Euclidean JT wormholes \cite{Penington:2019kki} (top) and corresponding microcanonical wormholes (bottom).}
    \label{fig:replica-wormhole-comparison}
\end{figure}

Let us compare our results to \cite{Penington:2019kki}. 
There, statistical averages for the R\'{e}nyi traces of a density
matrix were computed by defining the ensemble in terms of a dual replica wormhole computation in JT gravity. 
Here, in Section \ref{sec:higher-renyi}, we have done the same calculation, but our averages compute the result for a single, fixed microscopic theory where we have assumed our states are all typical in the microcanonical ensemble. 
One immediate simplification from this perspective is that our connected replica wormholes are just standard microcanonical wormholes, joined by topologically non-trivial wormholes. 
In our case, the bulk branes of \cite{Penington:2019kki} glue the distinct copies together seamlessly into a single connected boundary. We compare bulk geometries in Fig. \ref{fig:replica-wormhole-comparison}. 

\paragraph{Other ensembles.}

Although we find different bulk geometries from \cite{Penington:2019kki}, this should not be seen as 
evidence against its claims. 
Assuming our system is chaotic, we have argued that the right notion of a typical state in the microcanonical window is one drawn at Haar random. 
Adopting a different notion of typicality implies that more knowledge about the system is accessible to simple observers, such as more detailed information about the distribution of energy eigenstates in the microcanonical window, or other deviations from chaos. 
In these cases, there may very well be corrections to the moments of our distribution. 
Such corrections would require a modified gravitational interpretation (see for instance \cite{Balasubramanian2014, Saad2018, Cotler2016} and references therein).

Moreover, by restricting ourselves to states within a narrow microcanonical band, we have avoided questions about the dependence of the moments on energy differences, as displayed in the fuller version of ETH presented in Eq.\ \eqref{eq:ETH-correlator}. 
Understanding the relevant bulk geometries for these more general cases, and their relation to previous work, is of obvious interest.
What we wish to emphasize is that an understanding of the correct ensemble and notion of typical states need not come from averaging over theories, but can arise 
from a clearer operational understanding of how to integrate out microscopic splittings in an effective theory. 

 \paragraph{Quenched and annealed averages.}
In Sec. \ref{sec:second-renyi}, we calculated R\'{e}nyi entropies for highly entangled microcanonical density matrices.
This led to the appearance of correlated disorder between replicas, 
and hence a quenched average from the statistical perspective.
In this case, the connected contribution could compete with the disconnected part.

In contrast, the product of partition functions in Sec. \ref{sec:product-partition-functions} involved sums over microscopic states which were uncorrelated between copies.
A connected contribution arose from the cancellation of random phases, but this was always exponentially suppressed compared to the disconnected part, because there were no correlations.
This is an annealed average over disorder in the statistical picture, and a similar average arises when calculating R\'{e}nyi entropies for near-product density matrices. 
See \cite{Penington:2019kki} for a related discussion. 



 \paragraph{EFT and the RG.}
 
Here, we have used a notion of coarse-graining somewhat different from the standard Wilsonian perspective of integrating out high-energy degrees of freedom. 
It would be illuminating to directly relate the integrating out of microscopic splittings at high energy, as used in the ETH, to 
other approaches to renormalization and coarse-graining that have been studied in the holography and field theory literature, e.g. \cite{Calzetta:2008iqa,Bao:2019ghe,Lashkari:2018nsl,Kabernik:2019jko,BlumeKohout:2008zza,Almheiri:2014lwa}.

 \paragraph{Completing the EFT.}
 
 We have employed a 
 fairly skeletal
 notion of effective field theory in this letter, restricting our consideration to semiclassical bulk saddles but leaving the relation to the full gravitational effective field theory (including states with large deviations from the mean) unclear.
 
 Relatedly, we have concentrated on short-time physics, when states in the Schr\"{o}dinger picture have not had the opportunity to explore atypical corners of Hilbert space. 
 Many authors have explored the connection between ETH and gravitational saddles in this late time regime \cite{Maldacena:2001kr,Barbon:2003aq,Barbon:2004ce,Barbon:2014rma}, and it would be interesting to relate the present letter to this earlier work.

\begin{acknowledgments}
We thank Mark Van Raamsdonk for useful discussions. 
DW is supported by a UBC IDF Fellowship. 
We are supported by a discovery grant from NSERC of Canada, by the Simons Foundation, and by readers like you.

\end{acknowledgments}

\bibliography{refs}

\appendix
      
\section{Haar integrals}\label{app:unitary-integrals}

\subsection{Moments of typical states}\label{sec:typicality}

In this appendix, we will calculate typical quantities in the Haar-random microcanonical ensemble.
Consider the group of unitary matrices $\mathrm{U}(\He)$ over the microcanonical Hilbert space $\He$.
We are interested in moments of correlators of the form
\begin{align}
    \prod_{m=1}^n \bra{\psi_{i_m}}\calO_{a_m}\ket{\psi_{j_m}},
    \label{eq:gen-corr},
\end{align}
where the $\ket{\psi}$ are random.
To select such a typical state, we apply a Haar-random unitary $U \in \mathrm{U}(\He)$  to an arbitrary reference state $\ket{\psi_0}\in \He$.

To perform the ensemble average of (\ref{eq:gen-corr}), we set $\ket{\psi_{i_m}} = U_{(i_m)}\ket{\psi_0}$ and $ \ket{\psi_{j_m}}=U_{(j_m)}\ket{\psi_0}$, and integrate using the group-invariant measure over the choice of unitary operator.
To simplify the calculation, we insert a resolution of the (microcanonical) identity on either side of the operators $\calO_{a_m}$ in any convenient basis (labeled by $k_m$ and $l_m$):
\begin{align}
    \bra{\psi_{i_m}}& \calO_{a_m}\ket{\psi_{j_m}} \nonumber \\ & = \sum_{k_m, l_m}\bra{\psi_0}U^\dagger_{(i_m)}\ket{k_m}\bra{k_m} \calO_{a_m}\ket{l_m}\bra{l_m}U_{(j_m)}\ket{\psi_0} \nonumber\\
    & = \sum_{k_m,l_m}(\calO_{a_m})_{k_ml_m} U^\dagger_{(i_m)0k_m}U_{(j_m)l_m0} .\nonumber
\end{align}
Thus, we have
\begin{align}
 &\overline{\prod_{m=1}^n \bra{\psi_{i_m}}\calO_{a_m}\ket{\psi_{j_m}}}  = \prod_{m=1}^{n} \sum_{k_m,l_m} (\calO_{a_m})_{k_ml_m}  \nonumber\\ & \qquad \qquad \times \int  \mathrm{d}U_{(i_m)}\,\mathrm{d}U_{(j_m)}\, U^\dagger_{(i_m)0k_m}U_{(j_m)l_m0}.\nonumber
\end{align}
Symmetry under permutation of the indices $i_m, j_m$ dictates the form of the final answer:
\begin{align}
    & \overline{\prod_{m=1}^n \bra{\psi_{i_m}}\calO_{a_m}\ket{\psi_{j_m}}}= \label{eq:free-indices}\\
    & \quad = e^{-nS} N(\vec{i})\prod_{m=1}^{n} \sum_{\sigma \in S_n}\sum_{k_m,l_m}(\calO_{a_m})_{k_ml_m}  \delta_{i_mj_{\sigma(m)}}   \delta_{k_m l_{\sigma(m)}} \nonumber
\end{align}
for a normalization factor 
\begin{equation}
   N(\vec{i}) = e^{nS}\left(\prod_{l=1}^L (e^{S})_{{q_l}}\right)^{-1} \, .
\end{equation}
Here $(\cdot)_n$ is a rising Pochhammer symbol. 
We have also partitioned the set of states $\psi_{i_m}$ into component blocks of identical states. The sizes of these blocks are labeled $q_l$. 
This normalizaion can be found by choosing $\calO$ to be the identity. 


\subsection{Sums of typical states}

We can compute the ``empirical" microcanonical average by summing indices $i_{m}, j_m$ over $e^K$ random states.
Since the higher moments are exponentially suppressed, the central limit theorem gives
\begin{equation}
\sum_{i_m,j_m=1}^{e^K} \prod_{m=1}^n \bra{\psi_{i_m}}\calO_{a_m}\ket{\psi_{j_m}} \approx \sum_{i_m,j_m=1}^{e^K} \overline{\prod_{m=1}^n \bra{\psi_{i_m}}\calO_{a_m}\ket{\psi_{j_m}}}.\nonumber
\end{equation}
Let us calculate the RHS using the results of the previous subsection.
As before, symmetry implies
\begin{align}
& \sum_{i_m,j_m=1}^{e^K} \overline{\prod_{m=1}^n \bra{\psi_{i_m}}\calO_{a_m}\ket{\psi_{j_m}}} = \label{eq:full}\\
& \qquad \sum_{i_m,j_m}N(\vec{i})^{-1}\prod_{m=1}^{n} \sum_{\sigma \in S_n}(\calO_{a_m})_{k_ml_m}  \delta_{i_mj_{\sigma(m)}} \delta^{k_m}_{ l_{\sigma(m)}}.\nonumber
\end{align}

\subsection{Leading trace terms}

The equation \eqref{eq:full} can be rewritten using the structure of the permutations $\sigma$.
Suppose we have a cycle decomposition $\sigma = C_1 \cdots C_{c(\sigma)}$, where the cycle $C_i = (C_{i1}, \ldots, C_{id_i})$ has length $d_i$, including trivial cycles with $d_i = 1$.
For simplicity, we will focus on the `leading trace' terms, where for each trace structure we only keep the leading term.
After some algebra,
we obtain
\begin{align}
\sum_{i_m,j_m=1}^{e^K}& \overline{\prod_{m=1}^n \bra{\psi_{i_m}}\calO_{a_m}\ket{\psi_{j_m}}}\approx  \\ & \quad 
e^{-n(S-K)} \sum_{\sigma \in S_n} \prod_{i=1}^{c(\sigma)}\tr{\prod_{j=1}^{d_i}\calO_{C_{ij}}}.\nonumber
\end{align}

We can project onto a specific ordering of our states $j_m = \tau(i_m)$ by including a product of Kronecker deltas $\delta_{j_m\tau(i_m)}$ in the summation.
If such a term is already enforced by the permutation, it is absorbed, but each remaining delta reduces the number of terms by a factor $e^K$.
We therefore reduce by a total factor $\exp[Kd(\sigma, \tau)]$, where $d(\sigma, \tau)$ is the smallest number of transpositions required to change $\sigma$ into $\tau$, also known as the Cayley distance.
Explicitly, the leading trace contribution is
\begin{align}
&\sum_{i_m=1}^{e^K}\overline{\prod_{m=1}^n \bra{\psi_{i_m}}\calO_{a_m}\ket{\psi_{\tau(i_m)}}} \approx \label{eq:ident-perm} \\ & \qquad 
e^{-n(S-K)} \sum_{\sigma \in S_n} e^{- d(\sigma, \tau)K}\prod_{i=1}^{c(\sigma)}\tr{\prod_{j=1}^{d_i}\calO_{C_{ij}}}.\nonumber
\end{align}
Setting $\calO_a = \mathbb{I}$, operator traces give $e^{c(\sigma)S}$ and hence
\begin{align}
&e^{-nK}\sum_{i_m=1}^{e^K}\overline{\prod_{m=1}^n \langle \psi_{i_m}\ket{\psi_{\tau(i_m)}}}  \approx \label{eq:partition}\\ 
 & \qquad \qquad \qquad \sum_{\sigma \in S_n} e^{- d(\sigma, \tau)K - d(\sigma, \text{id})S} \, , \nonumber
\end{align}
This is the trace of a theory in an entangled random state $\rho_K = e^{-K}\sum_i \ket{\psi_i}\bra{\psi_i}$, on $n$ replicas identified via $\tau$.

Higher R\'{e}nyi entropies are evaluated using the cyclic identification $\sigma_n = (1\,2\, \ldots \, n)$. 
This leads to the expression for the leading trace:
\begin{align}
S_n(\rho_K) \approx \frac{1}{1-n} \sum_{\sigma \in S_n} e^{- d(\sigma, \sigma_n)K - d(\sigma, \text{id})S}\,.
\end{align}

\end{document}